# Dark angular momentum of the galaxy

Angelo Tartaglia

e-mail: angelo.tartaglia@polito.it
DISAT, Politecnico, Corso Duca degli Abruzzi 24, 10129 Torino, Italy

**Abstract:**

This paper proposes a strategy for detecting the presence of a gravito-magnetic field due to the rotation of the galactic dark halo. Visible matter in galaxies rotates and dark matter, supposed to form a halo incorporating barionic matter, rotates also, since it interacts gravitationally with the rest. Pursuing the same line of reasoning, dark matter should produce all gravitational effects predicted by general relativity, including a gravito-magnetic field. I discuss a possible strategy for measuring that field. The idea recovers the old Sagnac effect and proposes to use a triangle having three Lagrange points of the Sun-Earth pair at its vertices. The asymmetry in the times of flight along the loop in opposite directions is proportional to the gravito-magnetic galactic field.

# Introduction

Dark matter (DM) is one of the official ingredients of the standard cosmological model in its ΛCDM version. The question I would like to address is: independently from the numerous theoretical and experimental efforts to clarify what the nature of DM is, has the phenomenology at the scale of the Milky Way been fully explored in the framework of the standard paradigm of General Relativity (GR)? Indeed, if our galaxy is immersed in a DM halo with which visible matter interacts gravitationally, it is reasonable to expect that the dark halo too rotates as visible matter does. We then expect also that all related consequences of GR for rotating sources are produced.

In a weak field approach, we may speak of a gravito-electric component of the interaction, responsible for the ordinary gravitational pull, and a gravito-magnetic (GM) component producing a drag on moving masses. The question is: if GR is correct and DM exists, how could we detect the presence of a galactic GM field?

# The space-time geometry of a rotating mass distribution

Steady rotation implies an axial symmetry in space and, treating the galactic dark halo as a perfect fluid, we may use cylindrical space coordinates and write [1]:

$$ds^2 = e^{2U}(cdt + A\rho d\phi)^2 - e^{-2U}\left[e^{2k}(d\rho^2 + dz^2) + W^2\rho^2 d\phi^2\right] \quad (1)$$

The dimensionless functions $U$, $A$, $k$ and $W$ depend on $\rho$ (distance from the rotation axis, in flat space conditions) and $z$ (measured parallel to the axis) only. Space-time is assumed to be asymptotically flat. The origin of space coordinates is in the centre of the mass distribution; the frame is non rotating and at rest with respect to "fixed stars".

In weak field conditions, it is possible to interpret the off-diagonal time-space terms of the metric tensor, divided by the time-time term, as the components of a three-vector $\vec{h}$ acting as the vector potential of a gravito-magnetic field $\vec{B}_g$. It is:

$$\vec{B}_g = \vec{\nabla} \wedge \vec{h} \quad (2)$$

In the case of (1) we see that there is only one non-zero component of $\vec{h}$, namely:

$$h_\phi = \rho A \quad (3)$$

Consequently the only non-zero components of $\vec{B}_g$ are $B_{g\rho}$ and $B_{gz}$, however if we use Eq. (1) for the space-time of the Milky Way we may also include a symmetry with respect to the galactic plane, i.e. $g_{\mu\nu}(z) = g_{\mu\nu}(-z)$. This implies that on the plane $B_{g\rho}\big|_{z=0} = 0$. Adopting the viewpoint of a laboratory in the solar system we must transform the line element (1) by means of a boost at the peripheral speed of the Sun in the galaxy, $V$. The final result we are interested in concerns the form of $g_{00}$ and $g_{0i}$ in the laboratory frame, whence $h_i$, then $\vec{B}_g$ is obtained.

The calculation is lengthy but straightforward. Let us assume that the galactic trajectory of the Sun is a circle (radius $R$) and remember that $U$, $k$, $A$ << 1 and $W \sim 1$; it is also $V/c \sim 10^{-3}$. Of course the relative size of one term with respect to the others depends on the equation of state of the dark halo in the bulk. I can try a guess, having in mind the solutions in vacuo around a massive spinning body, with $U$, $k \sim 10^{-9}$, $A < 10^{-12}$ and $|W-1| < 10^{-12}$. The approximated calculation, up to terms linear in $A$, provides the only non-zero component of $\vec{B}_g$:

$$B_{gz} = \frac{\partial A}{\partial \rho} + \frac{A}{R} + 4\frac{V}{c}\frac{\partial U}{\partial \rho} + 2\frac{V}{c}\frac{k}{R}\bigg|_{z=0} \tag{4}$$

## Asymmetric light propagation

In the general case and for arbitrary coordinates we may solve the null line element for the elementary propagation time of electromagnetic (EM) signals [2]:

$$cdt = \frac{\sqrt{(g_{0i}dx^i)^2 - g_{00}g_{ij}dx^i dx^j}}{g_{00}} - \frac{g_{0i}}{g_{00}}dx^i \tag{5}$$

The sign in front of the square root has been chosen in order to insure propagation to the future ($dt > 0$).

If now we consider the time element along the same elementary space path for travel in opposite directions ($dx^i > 0$ in one case, $dx^i < 0$ in the other) and subtract a result from the other, then integrate the difference over a closed path, we get:

$$|\delta t| = \frac{2}{c}\oint \vec{h} \cdot d\vec{l} \tag{6}$$

Of course the integral is evaluated so that the result is positive; $d\vec{l}$ is the line element along the path. Pursuing the three-dimensional notation, applying Stoke's theorem, referring to metric (1), using a loop in the galactic plane and recalling formulae (2) and (3), we have:

$$|\delta t| = \frac{2}{c}\int_S \vec{B}_g \cdot \hat{n} dS \tag{7}$$

The flux integral is evaluated over the surface $S$ enclosed by the physical loop along which EM signals travel; $\hat{n}$ is the unit three-vector perpendicular to the surface element.

In the chosen reference frame the result depends both on the kinematical rotation with respect to "fixed stars" (classical Sagnac effect [3]) and on the GM field.

Recalling Eq. (4) we write explicitly:

$$|\delta t| \cong \frac{2}{c}\int \left(\frac{\partial A}{\partial \rho} + \frac{A}{R} + 4\frac{V}{c}\frac{\partial U}{\partial \rho} + 2\frac{V}{c}\frac{k}{R}\right)\bigg|_{z=0} dS \tag{8}$$

For experiments at the scale of the inner solar system we may expect changes in $B_{gz}$ at most of the order of

$$\delta B_{gz} \approx \pm \frac{\partial B_{gz}}{\partial \rho}\bigg|_{z=0, \rho=R} a \tag{9}$$

Length $a$ is of the order of the Astronomic Unit.

A reasonable guess for estimating the relative change of $B_{gz}$ across the inner solar system is of the order of $a/R$. The number is $\sim 6\times10^{-10}$ and the practical conclusion is that we may assume $B_{gz}$ as being constant, so that:

$$|\delta t| \cong \frac{2}{c}\left(\frac{\partial A}{\partial \rho} + \frac{A}{R} + 4\frac{V}{c}\frac{\partial U}{\partial \rho} + 2\frac{V}{c}\frac{k}{R}\right)\Bigg|_{\rho=R, z=0} S \quad (10)$$

## LAGRANGE

In order to implement the measurement outlined in the previous section we may think to the LAGRANGE proposal [4]. The idea is to locate EM signals emitters or transponders in the Lagrange (*L*) points of the Sun-Earth pair. The advantages of such configuration are that: a) polygons having the *L* points at the vertices are stable in time as for size and shape; b) the *L* points move around the barycentre of the Sun-Earth together with the Earth, so being at rest with respect to our planet.

We may think for instance to a triangle defined by $L_2 - L_4 - L_5$ (the enumeration is the standard one for Lagrange points). Seen from a terrestrial observer that configuration is a closed circuit contained in the plane of the ecliptic. The total area is $\sim 10^{22}$ m²: this number should be used in Eq. (10) as *S*. The plane of the ecliptic does not coincide with the galactic plane but the angle between the two is practically constant over times of years and corresponds to $\psi \sim 62.6°$. The inclination of the plane of the triangle provides a cosine factor $\sim 0.46$ in Eq. (10). For the LAGRANGE configuration, we rewrite (10) as:

$$\delta t \cong \frac{2}{c} B_{gz}\Big|_{\rho=R, z=0} S \cos\psi \quad (11)$$

Considering the numeric values, we see that the LAGRANGE experiment would be able to reveal a GM galactic field as small as:

$$B_{gz} \sim \frac{\delta t}{3\times10^{13}} \, m^{-1} \quad (12)$$

A picosecond time-of-flight asymmetry corresponds to a GM field $\sim 3\times10^{-26}$ m⁻¹. For comparison we have that the GM field on the surface of the Earth at the equator is [5]

$$B_\oplus = 2\frac{GJ}{c^3 R_\oplus^3} \cong 1.4\times10^{-22} \, m^{-1} \quad (17)$$

*J* is the angular momentum of the Earth. The GM field of the Sun at the terrestrial orbit is $B_{Sun} \sim 3\times10^{-27}$ m⁻¹ [4].

Of course we do not know how intense the GM field of the Milky Way can be as far as we have no solutions for the *U*, *k* and *A* functions of Eq. (10). *U* and *k* have to do with the gravitational potential, i.e. with the mass distribution, and *A* with the angular momentum, which means again the mass distribution and the rotation regime. I do not want to try here guesses of the equation of state of dark matter in the halo, but the orders of magnitude outlined above show that the sensitivities attained in the experiments aimed at revealing the terrestrial (or even solar) gravito-magnetism, if transposed to the LAGRANGE configuration, would also be sufficient to detect the presence of a weak, but not negligible, GM galactic field, if it exists.

### Further precisions
In formula (10) a factor *V/c* appears coming from a boost catching the rotation of the Sun around the centre of the Milky Way. However, in an experiment like LAGRANGE, we should include the rotation of the Earth (and the *L* points) around the Sun; this yields a variable contribution during the year, changing from parallel to antiparallel combinations of the velocities. In practice we expect an annual oscillation of approximately ±10% in the contributions coming from the third and fourth terms in brackets of formula (10). This seasonal change could help in spotting the GM galactic signal.

## Conclusion

The concept idea I have presented, when transformed into a real experiment, has to face a number of technical challenges both about the devices and about the modelling and control of the behaviour of spacecraft orbiting the *L* points. Existing technologies and the capacity to treat big data bases prove however to allow for very high sensitivities, as projects like LISA confirm. Furthermore, the *L* points are already the sites or targets for a large number of space missions, so that LAGRANGE could be considered as a comparatively low cost opportunity associated with other projects and would provide an interesting test for the presence and rotation of the dark galactic halo.

## References

[1] D. Kramer, H. Stephani, E. Herlt, M. McCallum, *Exact solutions of Einstein's field equations*, Cambridge University Press, Cambridge, p. 331 (2003)
[2] M.L. Ruggiero and A. Tartaglia, *Eur. Phys. J. Plus* **129**, p. 126 (2014)
[3] G. Sagnac, *C.R. Acad. Sci. Paris*, **157**, pp. 708-710 (1913)
[4] A. Tartaglia *et al.*, *Gen. Relativ. Gravit.*, **50**:9 (2018)
[5] F. Bosi *et al.*, *Phys. Rev. D*, **84**, 122002 (2011)